\documentclass[letter]{aa} 
\usepackage{graphicx}
\usepackage{txfonts}
\usepackage{lipsum}
\usepackage{subcaption}         
\usepackage{lscape}             
\usepackage{placeins}           

\makeatletter
\let\linenumbers\relax
\makeatother
\begin{document}

   \title{Identification of a dwarf galaxy stream in Gaia, and its possible association with the VPOS structure}


   \author{Hao Tian\inst{1,2},\thanks{tianhao@nao.cas.cn}
           Chao Liu\inst{1,2},
           Xiang-Xiang Xue\inst{1,2},
           Dongwei Fan\inst{1},
           Changqing Luo\inst{1},
           Jundan Nie\inst{1},
           Ming Yang\inst{1},
           Yujiao Yang\inst{3},
        \and Bo Zhang\inst{1}
        }

   \institute{National Astronomical Observatories, 
   Chinese Academy of Sciences, 
   Beijing 100101, People's Republic of China 
   \and Institute for Frontiers in Astronomy and Astrophysics, 
   Beijing Normal University, 
   Beijing, 102206, People's Republic of China
   \and School of Astronomy and Space Science, University of Chinese Academy of Sciences, Beijing, People’s Republic of China}

   \date{Received September 30, 20XX}

\abstract{
   Low surface density streams are important tracers to 
	study the formation and evolution of the Milky Way.
	  Using the accurate astrometric measurements 
	 from Gaia mission, we discover a low surface density stream in the north 
	 hemisphere, with length of $\sim110$ degree and width of $1.23$ kpc. The 
	vertical velocity dispersion perpendicular to the stream is around $22.4$ km s$^{-1}$.
	The spectral data of member candidate stars from LAMOST and DESI shows a
	large metallicity range from $-1.8$ to $-0.7$.
	Based on those properties we claim that
	the stream is originated from a dwarf galaxy.
	The median metallicity of $\mathrm{[Fe/H]}=-1.3$ indicates a massive
	dwarf galaxy origination with stellar  mass around $2.0\times10^7M_\odot$, which is 
	comparable with the Fornax dwarf galaxy and 
	 smaller than LMC/SMC and Sagittarius.
	 We also find the globular cluster Pyxis is highly associated with the stream
	 in the phase space $E-L_Z$ and metallicity.  
	 The massive progenitor also suggests that 
	  many dwarf galaxies, including  massive ones,
	  have been disrupted during their evolution orbiting the 
	  Milky Way and left with very low surface density structures. 
	  This is  important to understand the {\it missing satellites} problem. 
	  The orbit information of the stream shows 
	tight association between its progenitor and the 
	Vast POlar Structure (VPOS), which indicates that the satellites 
	fell into the Milky Way in groups, which brought many globular clusters into the Milky Way.}


   \titlerunning{Remnant of a massive dwarf galaxy}
   \authorrunning{Tian et al.}
   \maketitle
\section{Introduction}

During its formation, the Milky Way accreted 
and merged many satellites 
\citep{Johnston2008ApJ...689..936J}.
Those tidally stripped stellar systems, such as the
 globular clusters and dwarf galaxies, 
will left plenty of stars, which orbit the Milky Way
 with similar kinematic properties 
as their progenitors \citep{Helmi2020ARA&A..58..205H}. 
In this way, those stellar 
streams record the merging history
of the Milky Way. Since the discovery of the 
Sagittarius dwarf galaxy by 
\citet{Ibata1994Natur.370..194I} and its  tidal stream 
\citep{Ibata2001ApJ...547L.133I,Majewski2003ApJ...599.1082M}, 
around 200 stellar
streams have been found during the last two decades
 \citep{Mateu2018MNRAS.474.4112M,Ibata2024ApJ...967...89I}. 
Especially with the help
of  deep photometric surveys such as the SDSS
\citep{SDSS_York2000AJ....120.1579Y}, 
Pan-STARRS1 \citep{PS12016arXiv161205560C} and 
DES \citep{DES2016MNRAS.460.1270D}, 
plenty of cold streams have been discovered 
from the field stars with matched filter method 
\citep{Rockosi2002AJ....124..349R,Grillmair2006ApJ...639L..17G, 
Grillmair2006ApJ...643L..17G,Grillmair2006ApJ...645L..37G,
Grillmair2009ApJ...693.1118G,Grillmair2013ApJ...769L..23G,
Grillmair2014ApJ...790L..10G,Shipp2018ApJ...862..114S,
Grillmair2022ApJ...929...89G}. 
With precise proper motions provided by Gaia mission, 
\cite{Malhan2018MNRAS.477.4063M} developed a powerful 
method and discovered many 
thin weak streams, which were previously flooded in the field 
stars \citep{Malhan2019ApJ...886L...7M}. More recently,
\cite{Tian2024ApJ...965...10T} tried to remove the
 nearby field stars and discovered
a distant stream around  $25$ kpc. Moreover, combining the
 radial velocities (RV) 
from spectroscopic survey, more diffused substructures 
are discovered
in the phase space \citep{Koppelman2018ApJ...860L..11K,
YangCQ2019ApJ...880...65Y,Dodd2023A&A...670L...2D,Malhan2024ApJ...964..104M}.

What is more  interesting is that, many of those merged 
stellar systems did not fall into the Milky Way 
completely randomly, but 
in groups. 
Now we know that many globular clusters
 are associated with the Sagittarius system, such 
as Arp2, Berkeley 29, M 54, 
NGC 5634, Pal 12, Terzan 7, Terzan 8,
and Whiting-1 
\citep{Bellazzini2008AJ....136.1147B, 
Carballo-Bello2014MNRAS.445.2971C,
Dinescu2000AJ....120.1892D,
Martnez-Delgado2002ApJ...573L..19M,
Massari2019A&A...630L...4M,
NieJD2022ApJ...930...23N,
Vasiliev2019MNRAS.482.1525V}. 
With kinematic information,
\cite{Massari2019A&A...630L...4M} studied the 
formation of the globular clusters in the Milky Way.
They found $35\%$ globular clusters were associated
 with merging events during the formation 
of the Milky Way. For those dwarf galaxies, it 
shows that many of them are located 
in a very thin plane, which is almost perpendicular to
 the Galactic disk plane. That is also proved 
with the kinematic information. What is more, 
\cite{Pawlowski2012MNRAS.423.1109P}
showed that many young halo clusters and streams
 are also associated with a correlated population, which 
 is a vast structure composed by many dwarf galaxies 
 orbiting in a common polar plane,  known as the Vast POlar Structure (VPOS).
 However, the nature of this structure remains debated.  \cite{Riley2020MNRAS.494..983R} 
 examined the orbit normals of globular clusters and stellar streams and found no evidence of significant clustering.

To study the assemble history of the Milky Way with those streams,
 the spectroscopic observations are necessary. The spectroscopic
 surveys, such as LAMOST \citep{LAMOST2020arXiv200507210L},
   APOGEE \citep{Wilson2019PASP..131e5001W},
S$^5$ \citep{S52019MNRAS.490.3508L},
 \texttt{H3} \citep{H32019ApJ...883..107C} etc.,
 have been proposed
 to observe the member stars of 
those known streams to obtain the chemical 
information and radial velocities. 
Combining with the proper motions from Gaia,
 \cite{Naidu2020ApJ...901...48N}
reconstructed the formation history of the halo,
 and found that the Gaia-Sausage-Enceladus 
\citep{Helmi2018Natur.563...85H} dominated
the inner halo with galactocentric distance $r<25$ 
kpc, while the Sagittarius system dominated the outer 
part. What is more interesting is that, more than $95\%$ 
of their samples are linked to the 
merging substructures, which indicates
 that the halo has been  built 
by merged dwarf galaxies and the disk heating. 
Their results proved again that the 
outer halo was highly structured. 

As a results, a key question is that if all the satellites have been discovered.
\citet{Koposov2008ApJ...686..279K} showed that the luminosity function of the 
satellites  can be described by a single power law, ranging from $M_V=-2$
to the luminosity of the Large Magellanic Cloud. However, the number of currently discovered
satellites is much less than that predicted by the simulations, 
as known as the \emph{missing satellites} problem \citep{Klypin1999ApJ...516..530K}. 
One of the possible explanations is that part of those missing satellites 
 have been fully disrupted. In this way, there should be 
plenty of remnants, such as diffuse streams or clouds, left in the Milky Way.
The Orphan stream  is an example that many studies prefer its origination from
an ultra-faint dwarf galaxy \citep{Grillmair2006ApJ...645L..37G,Sales2008MNRAS.389.1391S}. 
It is important to discover those merging events, especially
those originated from merged dwarf galaxies  with low surface density. However, current 
methods have strong limitations on the discovery of those faint substructures.

To this end,   we focus on those fainter ones or those with low surface densities, which 
are possibly ignored or difficult to discover in previous work. We apply the similar 
method of \cite{Tian2024ApJ...965...10T} to Gaia DR3 and 
discover a new long stream.
This paper is constructed as follows. We briefly 
introduce the data selection and the method
in Section~\ref{sec:DataMethod}. The properties of
 the new stream are introduced 
in the Section~\ref{sec:NewStream}. Finally, we  
summarize the results in the Section~\ref{sec:Summary}.


\section{Data and Method} \label{sec:DataMethod}

In a small sky coverage, the stream member stars will have similar distance and velocities 
because all of them share unique orbit. Then those member stars will be in form of 
overdensities on the sky within corresponding proper motion ranges. The challenge is that 
low number of the member stars will make the stream flooded in the field stars.
Thus a very effecient way is needed to remove the contaminations
 of the field stars, especially
those disk stars, which are a few order higher than that of the halo in number density.
 This will significantly enhance the 
 signal of the streams \citep{Tian2024ApJ...965...10T}. 

Following \cite{Tian2024ApJ...965...10T}, we  
use the data from Gaia DR3 
to reveal the 
substructures in the halo. 
Here we briefly
introduce the steps.
Focusing on the substructures in the halo, 
especially those at high latitude,
we firstly select the stars with following criteria.
 \begin{enumerate}
 \item[1.]  $\omega<0.1$ mas
 \item[2.]  $\sigma_{\mu_\alpha^*}<0.2$ mas yr$^{-1}$ and $\sigma_{\mu_\delta}<0.2$ mas yr$^{-1}$
 \item[3.]  RUWE$<1.2$
 \end{enumerate}
 where $\omega$ is the parallax, $\sigma^*_{\alpha}$ and $\sigma_{\delta}$ are the uncertianties of the 
 proper motions, and RUWE is the Renormalised Unit Weight Error.

 \begin{figure}
  \centering
  \includegraphics[trim={6.7cm 0 7.cm 0.2cm},clip, width=0.45\textwidth]
  {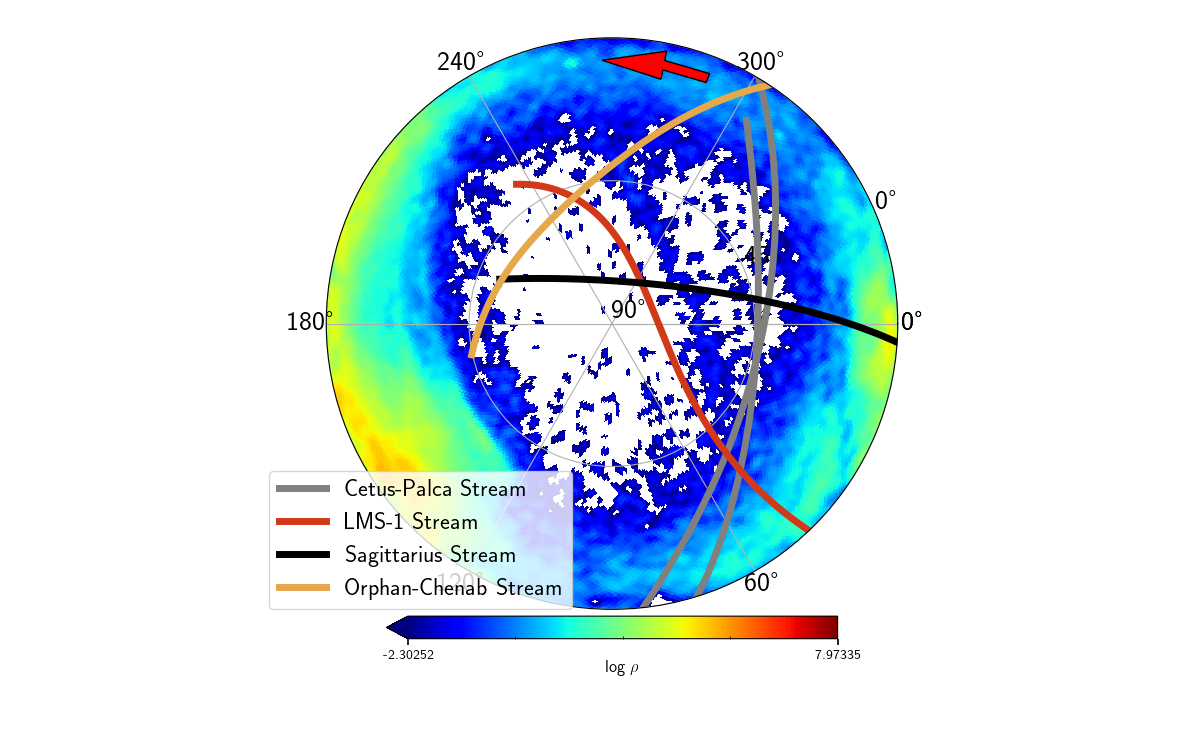}\\
  \vspace{-1.2cm}
   \caption{
   The density distribution  
   of the selected stars over the north semisphere.  
   {\bf The orbits of four streams
   are represented by the solid lines, which are provided by \cite{Mateu2018MNRAS.474.4112M} and \cite{ChangJiang2020ApJ...905..100C}.}
    The globular cluster 
   Pyxis is marked with red arrow with $(l,b)=(261^\circ.32, 7^\circ.00)$.
   }
  \label{fig:ra_dec}
\end{figure}
The first criterion removes most of the 
nearby stars. But a small fraction of distant stars with larger 
uncertainties of the parallax will be also removed. 
{\bf We do not correct the zeros-point of the parallax, 
which does not affect the final results, because parallax is only used removing the nearby stars.}
Assuming the parallax uncertainty is only
 relative to the magnitude, 
this will not affect the results, but slightly reduce
the number of the member stars, especilly those fainter ones. 
The second criterion ensures the 
proper measurements of the selected stars are accurate. 
This will remove most of the fainter stars, i.e. $G>19$.
In this way, those member stars from a 
unique stream will keep in a smaller range of proper 
motion. The last criterion is used to make sure that
most of the samples are single stars.

\section{The new stream} \label{sec:NewStream}
To figure out the substructure candidates, we slice those
selected stars  according to 
the proper motions in both $\alpha$ and $\delta$ direction, 
and collect the overdensities 
on the sky, which are of high probabilities to be a stream.

\subsection{Significance}
Based on the discussion above, here we report a significant 
new stellar stream in the north hemisphere with proper 
 motions $0.5<\mu_{\alpha^*}<1.5$ mas yr$^{-1}$ and
 $-0.5<\mu_{\delta}<0.5$ mas yr$^{-1}$. Figure~\ref{fig:ra_dec} shows
 the density distribution of all the stars satisfying 
 all the selection criteria including the constraints on the
  proper motions. The new stream candidate is located from top 
  $(l, b)=(270^\circ,45^\circ)$   to the 
  bottom right,  $(l, b)=(30^\circ,45^\circ)$, 
  which is around $110$ degree long. 
  Compared with the stellar stream library \texttt{galstreams} \citep{Mateu2018MNRAS.474.4112M}, 
  we find that this candidate does not overlap with any previously identified streams in the space,
  especially those originated from dwarf galaxies, 
  {\bf such as the Orphan Stream \citep{Belokurov2007ApJ...658..337B,Koposov2023MNRAS.521.4936K}, 
  the Cetus-Palca Stream
   \citep{Newberg2009ApJ...700L..61N,ChangJiang2020ApJ...905..100C,YuanZhen2022ApJ...930..103Y},
  the LMS-1 stream \citep{YuanZhen2020ApJ...898L..37Y,Malhan2021ApJ...920...51M} 
  as showed in Figure~\ref{fig:ra_dec}.}

 Following \cite{Tian2024ApJ...965...10T}, we also use the package 
  \texttt{gala}\footnote{https://gala.adrian.pw/en/latest/coordinates/greatcircle.html} \citep{gala}
  to rotate the coordinates from equatorial frame to a new one $(\phi_1, \phi_2)$
  making the stream along the equator in the new frame $(\phi_1, \phi_2)$, i.e. $\phi_2=0^\circ$.
  The stream is around $110$ degree long from $\phi_1=-45^\circ$ to
   $65^\circ$\footnote{The longitude range is directly
    relative to the selection of the two points} as represented by the shadow region in Figure~\ref{fig:phi1_phi2}.
    Fitting the latitude distribution with a Gaussian distribution, we find the 
  signal-to-noise ratio of the stream is around $19.6$. 
  The half width $(1\sigma_{\phi_2})$ is around  $2^\circ.57\pm0^\circ.52$.
  The background is around $H=12.54$ for each latitude bin with size of $1^\circ$, which 
  indicates the contamination of the field stars in each bin on average.
  Then we select all $290$ stars within
   $|\phi_2|<5^\circ$ (around $2\sigma$) as the member candiates,
   with around $H*10\,\mathrm{bins}=125$ stars from contamination. 
 
 \subsection{Geometric property}

 Cross-matching with the spectroscopic data, we 
 obtain spectra of 22 giant stars in total, $20$ stars from LAMOST
 DR9 and  $2$ from DESI early data release. 
 The top panel of Figure~\ref{fig:VVV} shows the distribution of
 the radial velocity ($\mathrm{RV}$) 
 versus the longitude $\phi_1$ 
 of 20 common stars, 
 the other two stars from LAMOST are not showed with $\mathrm{RV}$ around $-230$ km s$^{-1}$
 and metallicity lower than $-2$. All those $20$ common stars are selected as the member candidates 
 to study the properties of the stream. The metallicity of those common stars 
 has a larger variance, from $-1.8$ to $-0.7$, much larger than the uncertainties of $\sim0.1$ dex.
 What should be noticed is 
 that all those common stars are located on the right part of the stream 
 because of the sky coverages of LAMOST and DESI.

 First we estimate the metallicity of the stream
  with the median value of those $20$ member stars, 
 $\mathrm{[Fe/H]}=-1.3$. Then we try to fit the 
 distribution in CMD of all the member stars using the isochrone 
   with metallicity and age
   $(\mathrm{[Fe/H]},\mathrm{log_{10}}\,\tau)=(-1.3,10.08)$.
   With the package \texttt{dustmaps}
   \citep{Green2018JOSS....3..695M}, we correct the extinction 
    with the dust map obtained by 
   \citet{PlanckCollaboration2016A&A...586A.132P}
   and the extinction 
   coefficients of Gaia bands provided by
    \citet{Wang2019ApJ...877..116W}, i.e. 2.489, 3.161 and
     1.858 for $G$, $B_P$ and $R_P$ respectively.
   Then the distance  
   modulus is constrained as $\mathcal{DM}=17.19$
    ($d_{\mathrm{iso}}=27.42$ kpc). 
   Figure~\ref{fig:cmd} represents
   the distributions in CMD of the stars within 
   different longitude ranges from $-45^\circ$ to $65^\circ$ with step 
   of $22^\circ$. The selected 
    common stars with LAMOST and DESI are marked
     with red and magenta symbols, respectively. 
     From the middle and most right panel, there are two 
     stars significant offset from the fitted isochrone, which are highlighted 
     by the open circle in Figure~\ref{fig:VVV} and \ref{fig:cmd}. 
     Those two stars are not considered as member stars of the stream 
     in the following discussion.   
    What should be noticed is that there is no classical
    distance tracers in the members, such as RR Lyrae stars
    or blue horizontal branch stars. {\bf So here we
    adopt the distance $d_{\mathrm{iso}}=27.42$ kpc for all the member stars.}
    According to the position of the possible red horizontal branch 
    stars,  approximately located at
    $(B_P-R_P, G)=(0.8, 17.6)$, which is marked with 
    the horizontal dashed line in each panel,
     there is not significant gradient versus the 
     longitude $\phi_1$. In this way, we  firstly
     estimate the width of the stream by 
     $d_{\mathrm{iso}}*\sigma_{\phi_2}=1.23$ kpc. 

     The larger variance of the metallicity  and the
     width discussed above suggest that the stream possibly originates from
     a dwarf galaxy, rather than a globular cluster.

\subsection{Kinematic property}

{\bf Considering most of the member candidate stars without spectroscopic observation, we only
investigate the properties of the tangential velocities, which are perpendicular to the line of sight.}
The bottom panel of Figure~\ref{fig:VVV} shows the distribution of the 
tangential velocities after the correction of all the member candidate stars
with $G<18$. The shadow region represent the ranges caused by the proper motion 
selection. We find the  dispersion of the tangential 
velocity perpendicular to the stream is around $30.7$ km s$^{-1}$
for all those members with $G<18$.
The dispersion is an order higher than that of a cold stellar stream
 formed from a globular cluster, such as $2.1\pm0.3$ km s$^{-1}$ 
 for GD-1  stream \citep{Gialluca2021ApJ...911L..32G}.
 Considering the maximum uncertainties of proper motion,
  $0.2$ mas yr$^{-1}$, the contribution to the dispersion
  is $4.74*d*\sigma_\mu\sim16.5$ km s$^{-1}$. Assuming a median proper motion 
  of $1$ mas yr$^{-1}$ and relative distance uncertainty of $10\%$, the
   distance uncertainty will contribute $\sim13.0$ km s$^{-1}$.
  Then the instrinsic  dispersion can be estimated by 
  $(30.7^2-16.5^2-13.0^2)^{0.5}=22.4$ km s$^{-1}$,
  which is still an order higher than that of the 
  cold stream and almost all the globular 
  clusters\footnote{https://physics.mcmaster.ca/\textasciitilde harris/mwgc.dat} 
  \citep[2010 version]{Harris1996AJ....112.1487H}.
  The larger tangential velocity dispersion also suggests that
the stream is formed from a disrupted dwarf galaxy. 
Condisdering the stream is not orbiting a great circle on the sky, which will
enlarge the dispersion of all the member candidates, we also
check the dispersion in different longitude ranges as represented by the green 
symbols, we find the dispersion is still significant enough in each subsample.

\begin{figure*}
  \centering
  \includegraphics[trim={2cm 0 2cm 3cm},clip, 
  width=0.45\textwidth]{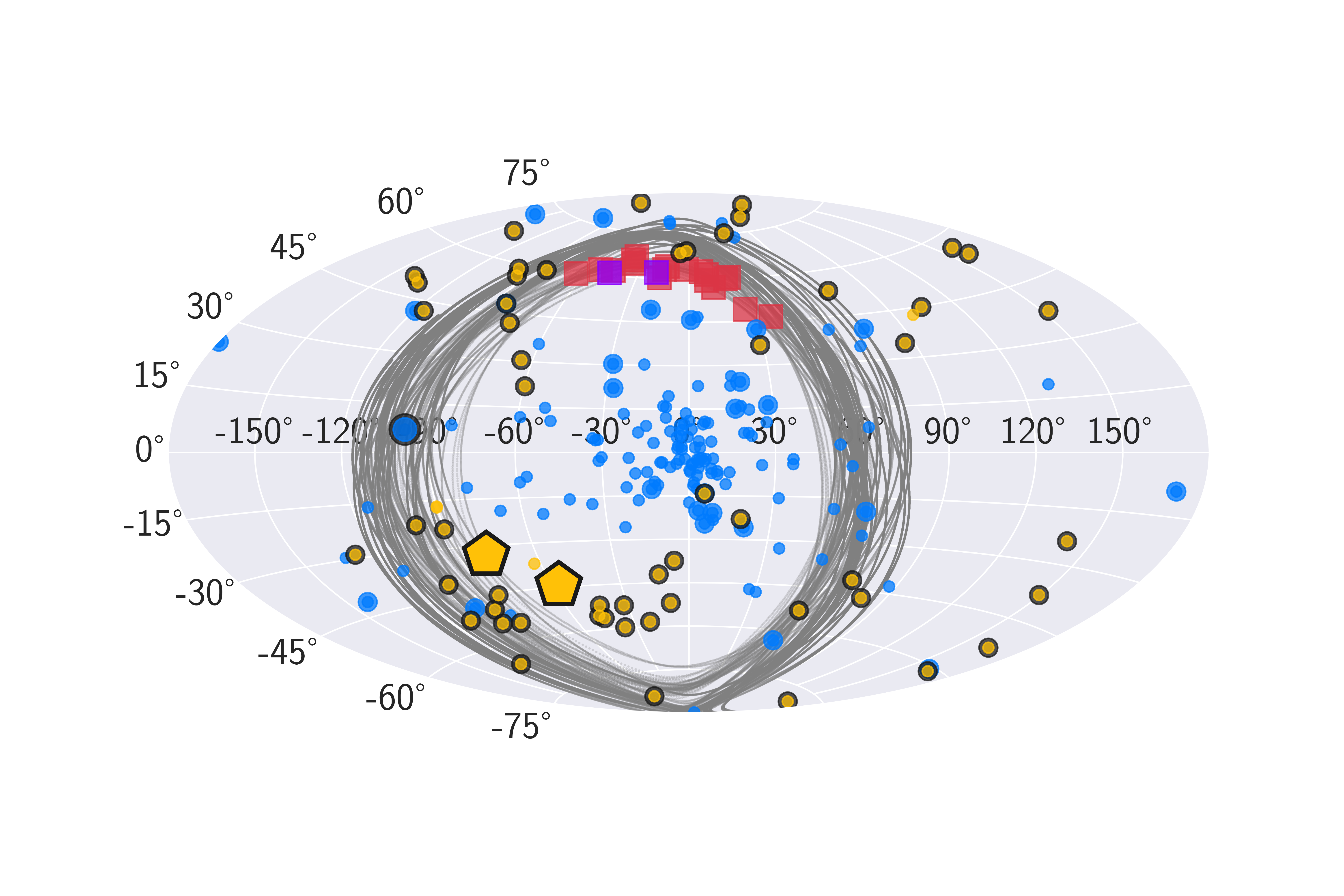}
  \includegraphics[trim={2cm 0 2cm 3cm},clip, 
  width=0.45\textwidth]{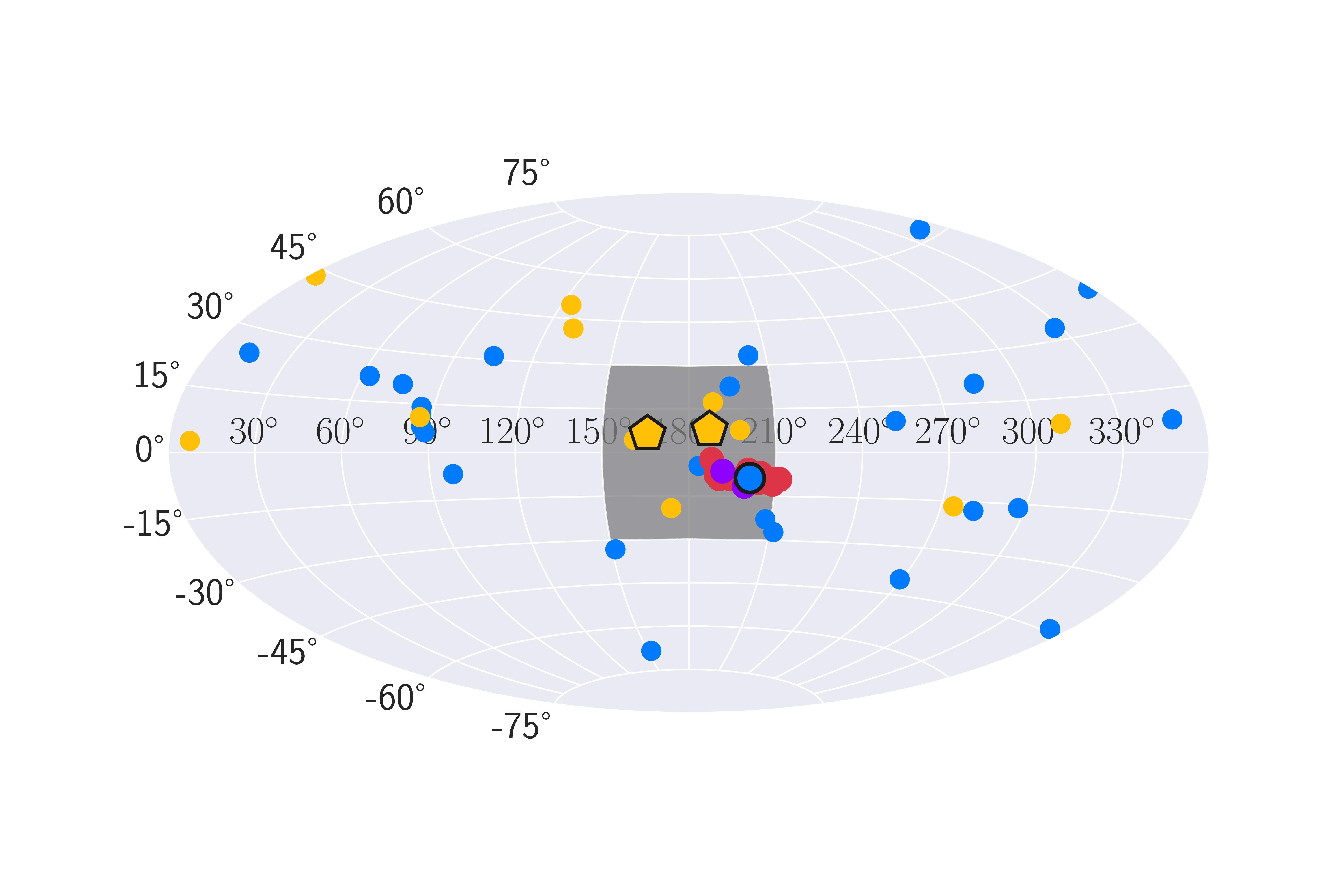}\\
  \vspace{-1.45cm}
   \caption{
  Left : Sky projection of those common stars, globular clusters and dwarf galaxies
  represented by the red/magenta squares, blue dots and yellow dots, respectivly.
  The globular cluster Pyxis and LMC/SMC are highlighted with larger blue dot and 
  yellow pentagons. All those globular clusters and dwarf galaxies with distance
  larger than 20 kpc are hightlighted with solid black edge.
  {\bf  The integrated orbits 
  of those stars with full 6D information  are represented by the gray lines.
   Right: The direction distribution of the  orbit normals
of those common stars, distant globular clusters and dwarf galaxies with full 6D information
are represented. The symbols are the same with that in the left panel. The shadow region
represents the direction of the normal of the VPOS
 with longitude between $150^\circ$ and $210^\circ$ and latitude between $-30^\circ$ and $30^\circ$.}
   }
  \label{fig:Orbit}
\end{figure*}

\begin{figure*}
  \centering	
  \includegraphics[width=0.4\textwidth]
	{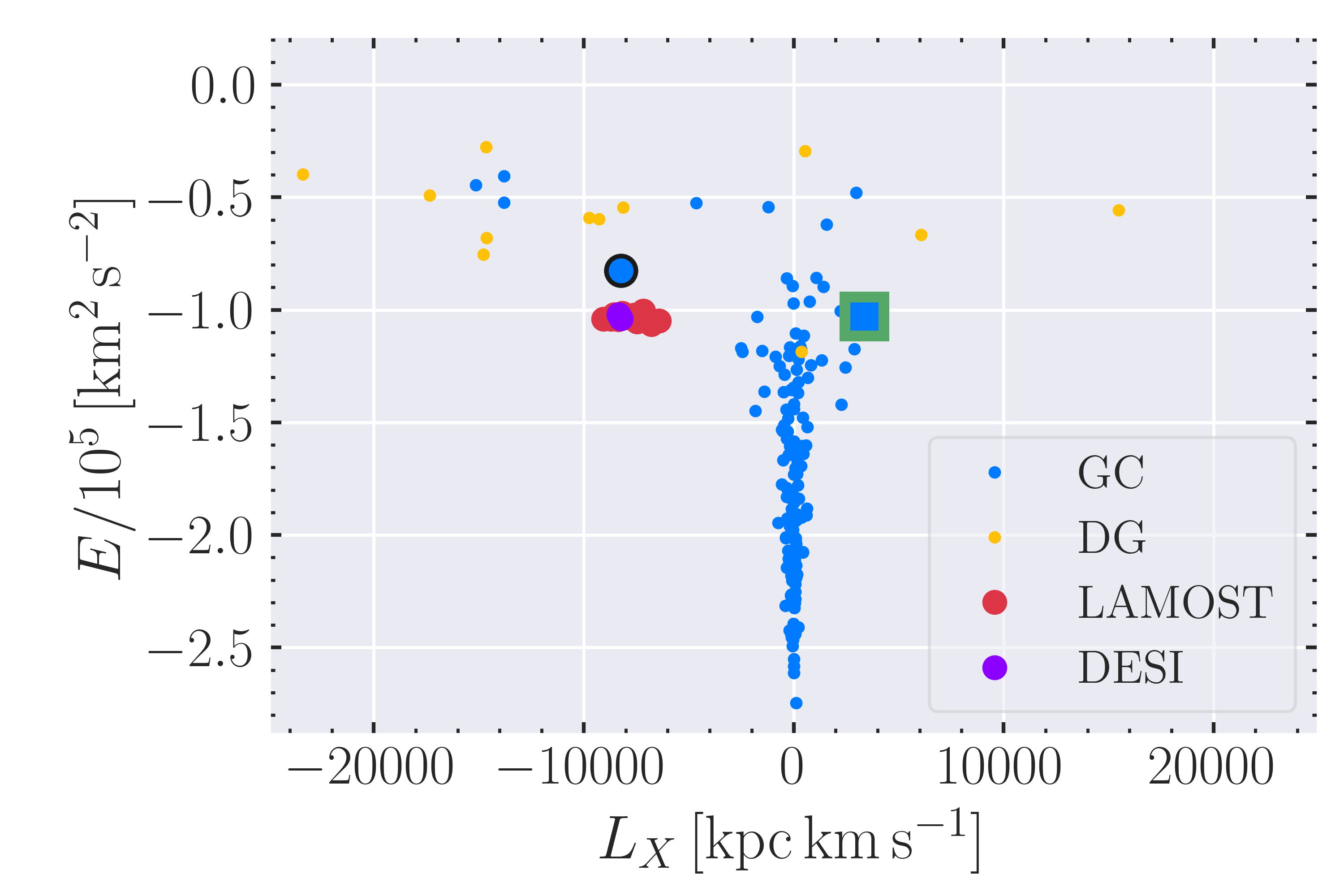} \hspace{0.75cm}
  \includegraphics[width=0.4\textwidth]
	{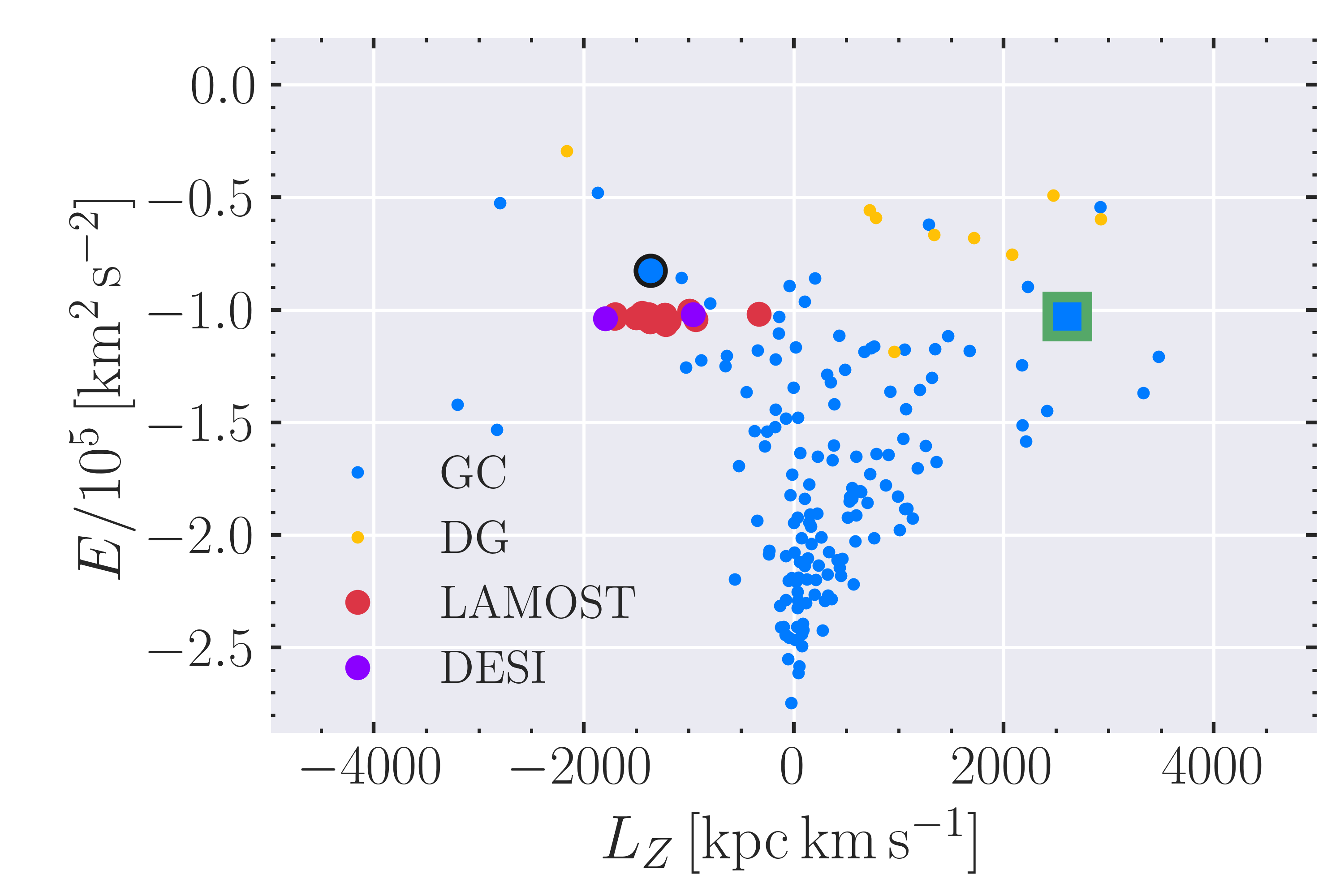}\\
  \vspace{-0.45cm}
   \caption{The distributions of the common stars, the globular clusters
	 and the dwarf galaxies are represented by the red/magenta larger dots, the blue and yellow dots
    in the phase spaces  $E$ versus $L_X$ (left) and $L_Z$ (right), respectively. {\bf The locations of Pyxis and
    NGC 5824 are highlighted by the larger blue dot and square, respectively.}
   }
  \label{Fig:ELx}
\end{figure*}

\subsection{Origination} 
The large width of $1.23$ kpc and the velocity dispersion of
 $22.4$ km s$^{-1}$ indicate
the stream is originated from a dwarf galaxy. 
The stellar mass of 
its progenitor dwarf galaxy can be estimated as $M_*\sim2.0\times10^{7}M_\odot$
with the universal metallicity-mass 
relation provided by \cite{Kirby2013ApJ...779..102K}, 
adopting the metallicity of $-1.3$.
That suggests a 
massive progenitor of dwarf galaxy, which 
is smaller than the LMC/SMC and the Sagittarius
 system, and comparable with that of the 
 Fornax spheroidal dwarf galaxy. 
It is important to note that this relation 
was derived from dwarf galaxies in the Local Group, with an RMS of 0.17. 
Meanwhile, the metallicity-mass relation evolves 
such that higher-redshift galaxies of similar 
mass tend to be more metal-poor \citep{Erb2006ApJ...644..813E,Zahid2013ApJ...771L..19Z, Henry2013ApJ...776L..27H}.

To further investigate its origination, we integrate the orbit of those common
stars with full information forward and backward with 
 the package \texttt{AGAMA} \citep{Vasiliev2019MNRAS.482.1525V}
 and represent the orbits  in Figure~\ref{fig:Orbit} with gray lines.
 The potential of the Milky Way is assumed 
 to be  composed by three components, a Dehnen bulge, 
a Miyamoto-Nagai disk and an NFW halo. The masses for those 
three parts are $2.0\times10^{10}\,M_\odot$, 
$5.0\times10^{10}\,M_\odot$ and 
$5.5\times10^{11}\,M_\odot$, respectively. 
{\bf The scale radii are set
$1.0$ kpc, $3.0$ kpc and $15.0$ kpc, then the total mass enclosed within 20 kpc is around $2.1\times10^{11}M_{\odot}$. }
The eccentricity and pericenter distance are of $\sim0.7$ and $\sim24$ kpc, respectively. 
The positions of the globular clusters
 \citep{Vasiliev2019MNRAS.482.1525V} and
 dwarf galaxies \citep{Drlica-Wagner2020ApJ...893...47D}
 are also presented by the 
blue and yellow dots. The LMC and SMC are 
highlighted with larger yellow pentagons. 
{\bf From the sky distribution, we find that many dwarf
 galaxies and distant globular clusters are
located on the orbit of the stream as showed in the left panel of Figure~\ref{fig:Orbit}.}

This is also proved by the orbit direction distribution 
in the right panel of Figure~\ref{fig:Orbit}.
We can find two groups, the first one is located around $(0^\circ,90^\circ)$
where  many globular clusters associate with the Sagittarius Stream. 
The other one is around the direction $(0^\circ,180^\circ)$, where locates 
dwarf galaxies associated with the Vast POlar Structure (VPOS)
including the
  LMC, SMC, Carina, Draco, Fornax and 
  Ursa Minor \citep{Pawlowski2012MNRAS.423.1109P}.
 The orbit norm directions (direction of the angular momentum vector 
 $\vec{L}=(L_X,L_Y,L_Z )$)
of those common stars 
 are also located in the similar region around 
$(\phi,\theta)=(202^\circ.9, -8^\circ.9)$,
which indicates an association 
between the new stream and VPOS.


Checking the phase space distributions of those globular clusters and dwarf galaxies, 
we find the Pyxis globular cluster is highly associated
with the stream as showed in Figure~\ref{Fig:ELx}. 
First, it locates on the elongation of the stream.
The location of Pyxis is highlighted by the arrow in Figure~\ref{fig:ra_dec},
and by the black-edged blue dot around
 $(l,b)\sim(-99^\circ,7^\circ)$ in the left 
panel in Figure~\ref{fig:Orbit}.
What is more, 
the stream and Pyxis have similar angular momenta, i.e. 
the directions of the orbits.
Second, Pyxis has a similar metallicity $-1.2$ \citep{Harris1996AJ....112.1487H} 
and CMD distribution. Figure~\ref{fig:cmd} shows the CMD distribution of the 
members stars of the stream and the globular cluster Pyxis with black and blue
dots, respectively. The members of  Pyxis are shifted to the distance of the stream.
The extinction of Pyxis from \cite{Vasiliev2019MNRAS.482.1525V} is adopted. 
We find the positions of the red horizontal branch stars are quite consistent.
{\bf Meanwhile, we do not find any connection with the Cetus-Palca Stream, 
which is located mainly in the south semisphere and
tightly associated with the globular cluster NGC 5824 \citep{ChangJiang2020ApJ...905..100C}, 
which is highlighted with square in Figure~\ref{Fig:ELx}.
}

All the evidences above 
suggest that the stream and the globular cluster Pyxis
have the same origination, a dwarf galaxy with stellar
 mass of $2.0\times10^7M_\odot$.
The new stream and the globular 
cluster Pyxis, are a close analogy  
as the Sagittarius system or the $\omega$-Centauri system, 
composed of a stripped stream, and
at least one globular cluster and the progenitor dwarf galaxy. 
What is different is that the new stream
has a much larger pericenter distance than the 
Sagittarius and the  $\omega$-Centauri system.
That means  the tidal effect from the Milky Way
 is much smaller
than that to the  Sagittarius and the  $\omega$-Centauri system. 
What should be noticed is that Pyxis is  a quite
special globular cluster with very large half-light 
radius,  $r_{\mathrm{h}}=17.7$ pc \citep{Fritz2017ApJ...840...30F},
larger than that of almost all
the current discovered globular clusters in the Milky Way and
 close to the bottom limit of the dwarf
galaxies \citep{Drlica-Wagner2020ApJ...893...47D}. 
The large pericenter distance makes the globular cluster Pyxis survived 
from the tidal effect of the Milky Way during
 the disruption of the progenitor dwarf galaxy. 

According to the results from \cite{Hoyer2021MNRAS.507.3246H} 
based on the analysis of $601$ galaxies
in the local volume, the occupation of the nuclear star
 cluster in the galaxies with 
stellar mass $<10^{9.5}M_\odot$ can reach up to  $40\%$.
 More evidence is necessary 
for further studies on the relation between the globular
 cluster Pyxis and the progenitor 
dwarf galaxy, i.e. if Pyxis is the nuclear star cluster
or a satellite globular cluster of the progenitor dwarf galaxy,
or Pyxis is the remnant of the core of the dwarf galaxy, whose
outskirt has been completely stripped.
In any case, it is possible that
 the  progenitor dwarf galaxy of the stream has been 
 almost fully disrupted  like the $\omega$-Centauri system \citep{Ibata2019NatAs...3..667I} . 

 The discovery of this new stream suggests again that many globular clusters,
  especially those relatively metal rich ones,
  are not formed in the Milky Way, but in those 
  merged dwarf galaxies. Even more, 
  some of those globular clusters are the 
  nuclear star cluster of the stripped galaxies
  \citep{Zinn1993ASPC...48...38Z,Muratov2010ApJ...718.1266M,
  Renaud2017MNRAS.465.3622R,Drlica-Wagner2020ApJ...893...47D}.
 Meanwhile the discovery of the new stream 
 also indicate that many dwarf galaxies have been 
 stripped during the evolution of the Milky Way, even those massive ones. 
 This at least partly explains the 
 {\it missing satellites} problem.

\section{Summary} \label{sec:Summary}
 Focusing on the distant volumes, we remove majority
  of the nearby stars with parallax provided by 
  Gaia DR3. Slicing the proper motions in both 
  directions along $\alpha$ and $\delta$, we find 
  a $110$ degree long stellar stream with
   $0.5<\mu_{\alpha^*}<1.5$ and
 $-0.5<\mu_{\delta}<0.5$ mas yr$^{-1}$, which
  is around $\sim27.42$ kpc from the Sun. 
 Through the number fitting along the latitude
  $\phi_2$ perpendicular to the elongation of the stream
 with a Gaussian profile, we find the half width is 
 around $1.23$ kpc (1$\sigma$ in Gaussian).
 The tangential velocity dispersion perpendicular
  to the stream is around $22.4$ km s$^{-1}$. The
 larger width and velocity dispersion indicate an
  origination from a dwarf galaxy. With the 
 help of LAMOST and DESI, we obtain the full 
 information of $18$ stars of high probabilities of stream member. 
 The metallicity $\mathrm{[Fe/H]}=-1.3$ indicates
 that the stellar mass of its progenitor is 
 around $2.0\times10^7M_\odot$, which is much larger
 than the majority of the dwarf galaxies in the Milky Way, 
 next to Magellanic system, the Sagittarius 
 and Fornax dwarf galaxies. What is more,
 the orbits of those stars
 indicate that this stream is associated with the VPOS. 
 What is more important, we find the globular
  cluster Pyxis is tightly associated 
 with the new stream. Analogy with the 
 $\omega$-Centauri system, the progenitor dwarf
  galaxy should have been completely disrupted.  
 It is not clear that if Pyxis is 
 the nuclear star cluster or a satellite globular
  cluster of the progenitor dwarf galaxy. 
 This discovery proves the path 
 of the formation of the globular clusters in the Milky Way,
 that many of which are formed from the 
 merged dwarf galaxies.
 This new discovered stream also indicates that the missed 
 satellites, even massive ones, may have been disrupted during 
 the evolution. Tidal disruption is one of the solutions
 for the {\it missing satellites} problem. There should be
 many low surface density diffuse streams in the halo.


\begin{acknowledgements}
We thank the referee for those comments which greatly improved the manuscript.
This work is supported by National Key R\&D Program of 
China No. 2024YFA1611902 and the China Manned Space Project. 
X-X.X. acknowledges the support from CAS Project for
 Young Scientists in Basic Research Grant 
No. YSBR-062 and NSFC grants No. 11988101. 
D.F. acknowledges the support from National Natural 
Science Foundation of China with Grant No.12273077.
J.N. acknowledges the supports by the 
Beijing Natural Science Foundation (grants 
No.1232032), by the National Key R\&D Program 
of China (grants No. 2021YFA1600401,2021YFA1600400), 
by the Chinese National Natural Science Foundation 
(grants No. 12373019).
M.Y. is supported by the National Natural Science 
Foundation of China (Grant No. 12373048)
Y.Y. acknowledges the support from National Natural 
Science Foundation of China with Grant No.12203064.

Data resources are supported by China National 
Astronomical Data Center (NADC) and Chinese Virtual
 Observatory (China-VO). 
This work is supported by Astronomical Big Data Joint 
Research Center, co-founded by National Astronomical Observatories, 
Chinese Academy of Sciences and Alibaba Cloud.

This work has made use of data from 
the European Space Agency (ESA)
mission {\it Gaia} (\url{https://www.cosmos.esa.int/gaia}), 
processed by
the {\it Gaia} Data Processing and Analysis Consortium (DPAC,
\url{https://www.cosmos.esa.int/web/gaia/dpac/consortium}).
 Funding
for the DPAC has been provided by national institutions,
 in particular
the institutions participating in the {\it Gaia} 
Multilateral Agreement.
Guoshoujing Telescope (the Large Sky Area Multi-Object
 Fiber Spectroscopic Telescope LAMOST)
is a National Major Scientific Project built by the 
Chinese Academy of Sciences. Funding for 
the project has been provided by the National 
Development and Reform Commission. 
LAMOST is operated and managed by the National 
Astronomical Observatories, Chinese Academy of Sciences.
\end{acknowledgements}

\bibliographystyle{aa} 
\bibliography{main} 

\begin{appendix}

\onecolumn

\section{space distribution}
\begin{figure*}[h]
	  \includegraphics[ width=0.94\textwidth]
	  {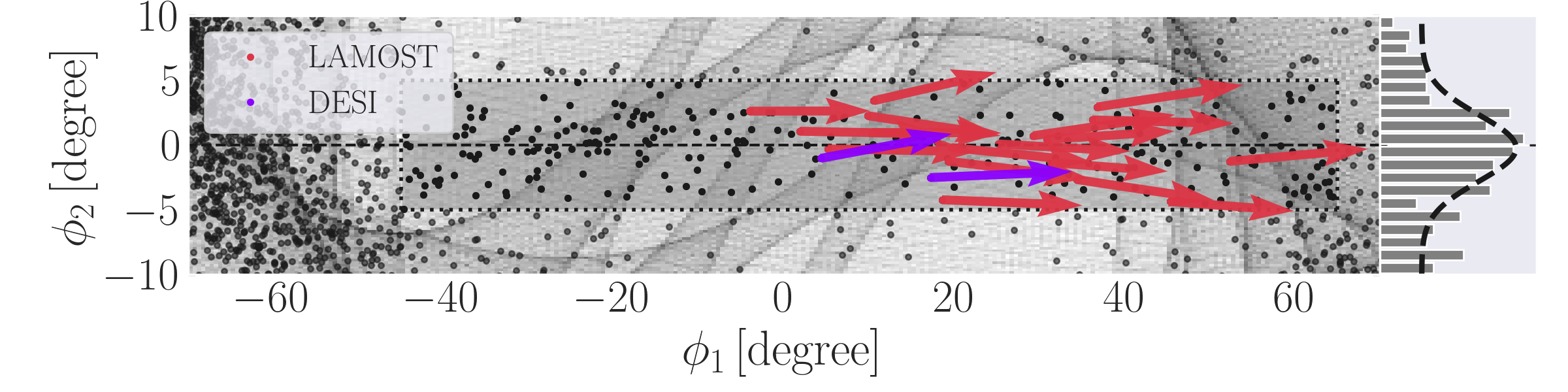}\\
	   \caption{The sky distribution of the stars in the new coordinates is showed 
	   in the left panel. The shadow region represents the coverage of the new stream
	   with $-45^\circ<\phi_1<65^\circ$ and $|\phi_2|<5^\circ$.  The stars with arrows 
	   represent the stars with RV from LAMOST (red) and DESI (magenta). The arrows
	   represent the tangential velocities with Solar motion corrected.
	   The right panel shows the number distribution of the stars along the latitude $\phi_2$ with
	   longitude $-45^\circ<\phi_1<65^\circ$. The dashed line represents the fitting results
	   with a Gaussian distribution.
	   }
	  \label{fig:phi1_phi2}
	\end{figure*}

The coordinates rotation is done with the package \texttt{gala}. The two coordinates 
on the stream  $(\alpha_1,\delta_1)=(169.6, -16.9)$
  and $(\alpha_2,\delta_2)=(216.9, 11.5)$ are adopted. The left panel in Figure~\ref{fig:phi1_phi2}
  shows the distribution of the stars in the new coordinate frame. The shadow region
  represents the selection for the member stars. The right panel represents the number distribution
  along the latitude $\phi_2$ with $-45^\circ<\phi_1<65^\circ$. The dashed line represents the fitting 
  results with the  latitude  dispersion of  $\sigma_{\phi_2}=2^\circ.57$ for 
  a Gaussian distribution $y=A*Gaussian(\bar{\phi_2},\sigma_{\phi_2})+H$. The background $H=12.54$.

\section{Distance}
\begin{figure*}[h]
    \centering
    \includegraphics[trim={0.0cm 0 0.2cm 0},clip, 
    height=0.34\textwidth]{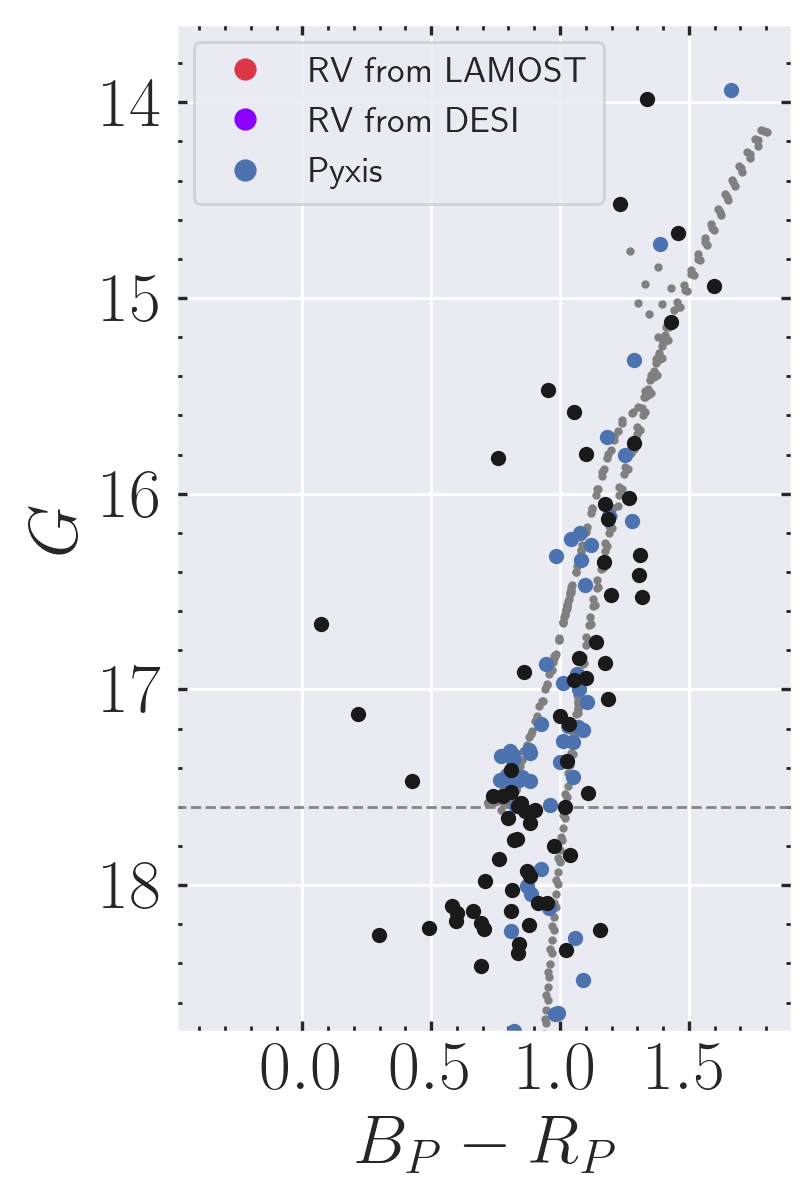}
    \includegraphics[trim={2.1cm 0 0.2cm 0},clip, 
    height=0.34\textwidth]{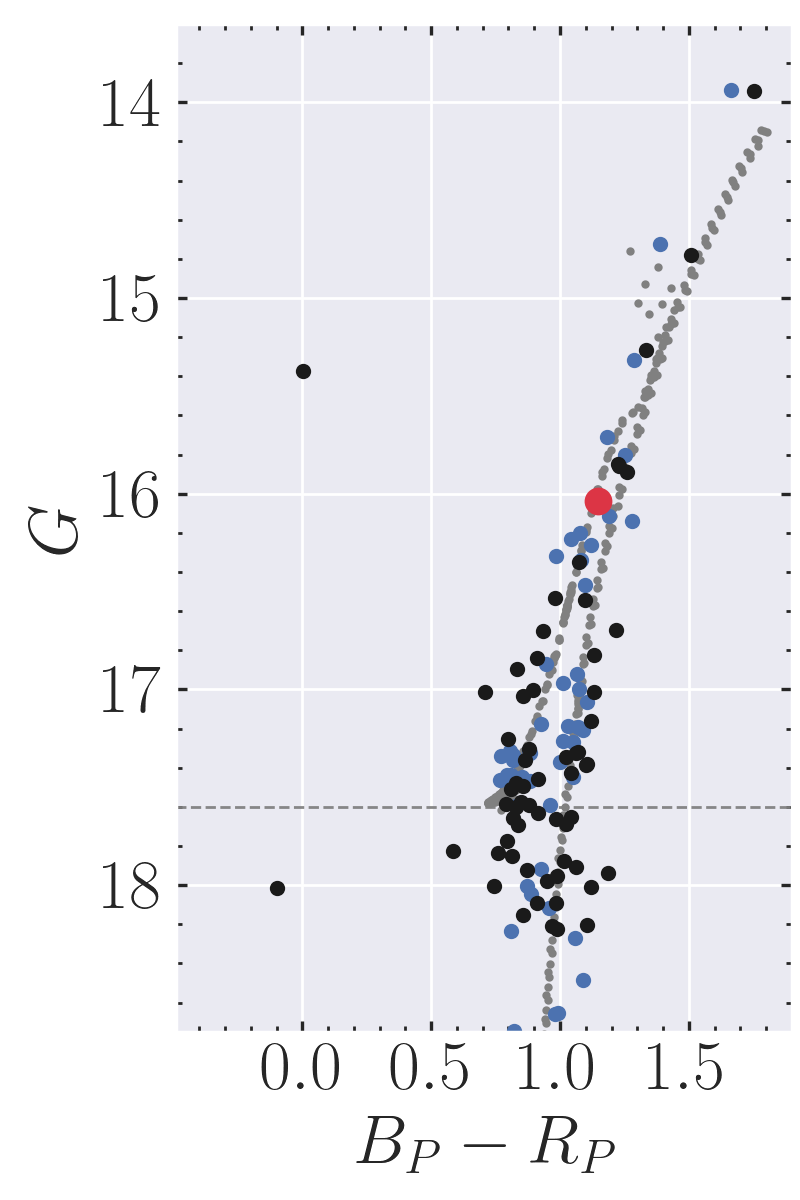}
    \includegraphics[trim={2.1cm 0 0.2cm 0},clip, 
    height=0.34\textwidth]{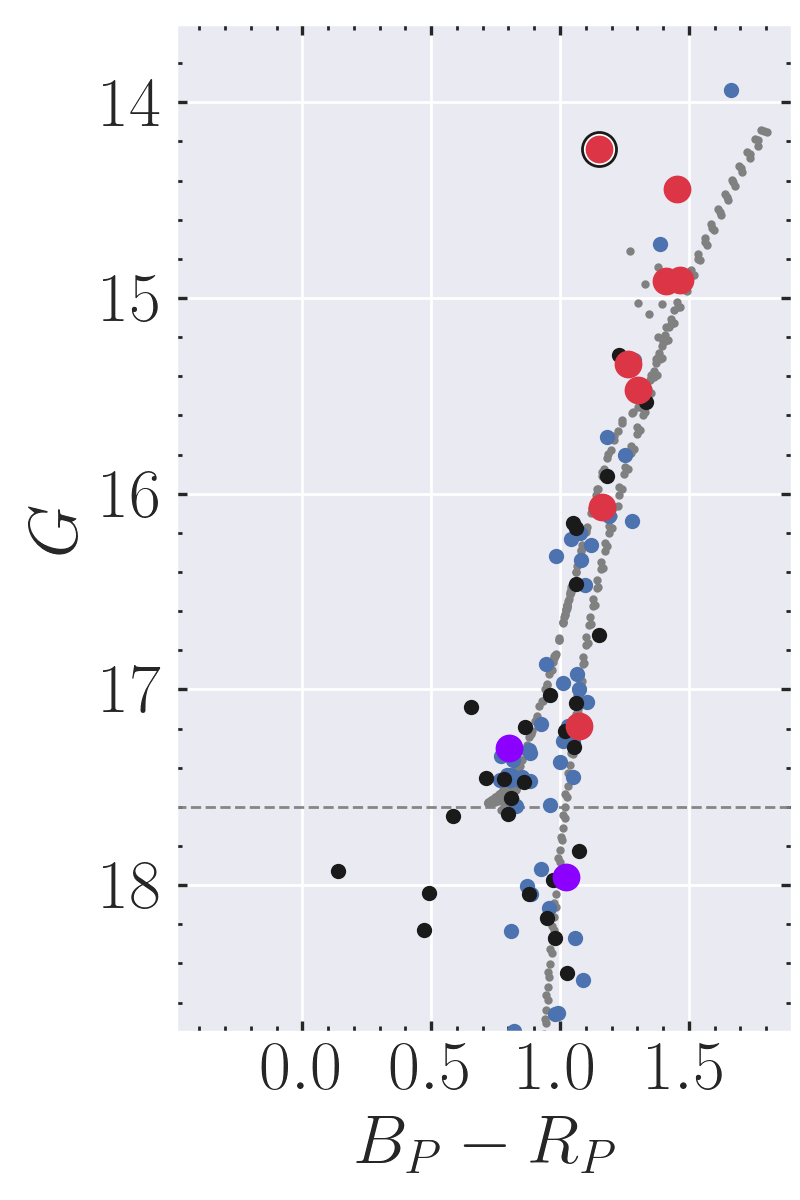}
    \includegraphics[trim={2.1cm 0 0.2cm 0},clip, 
    height=0.34\textwidth]{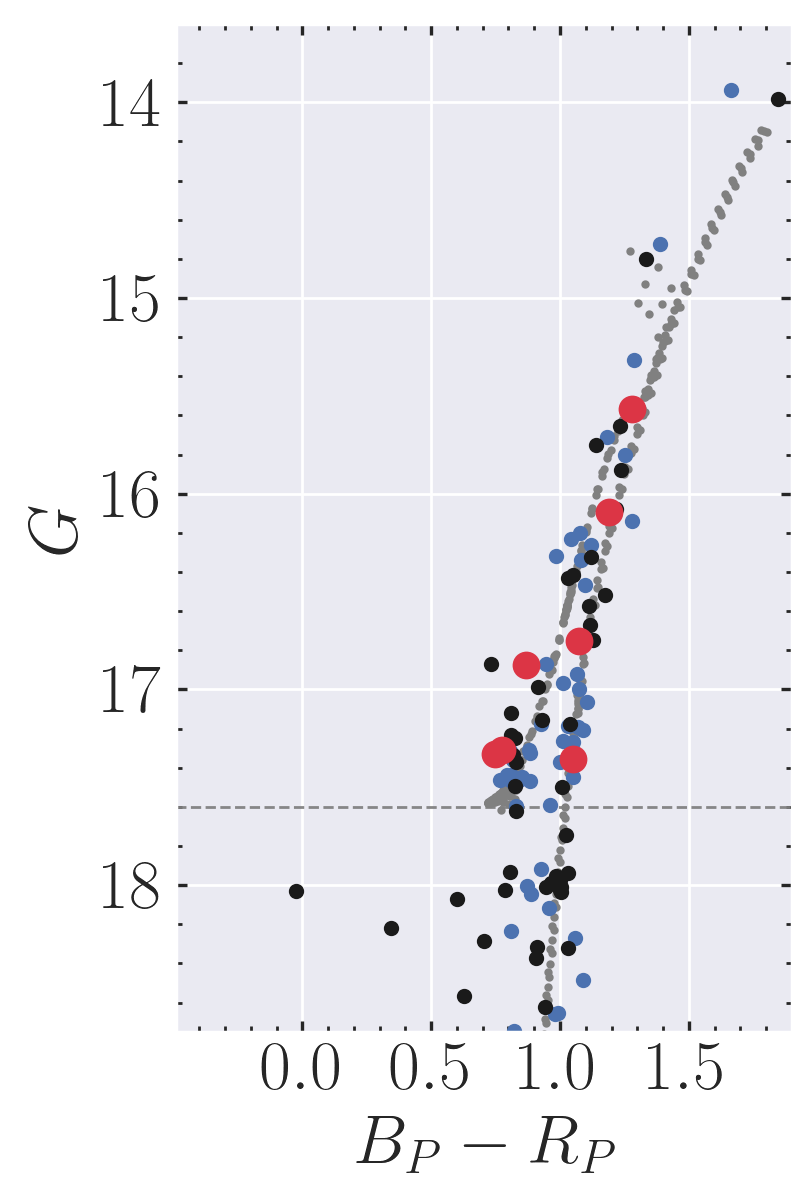}
    \includegraphics[trim={2.1cm 0 0.2cm 0},clip, 
    height=0.34\textwidth]{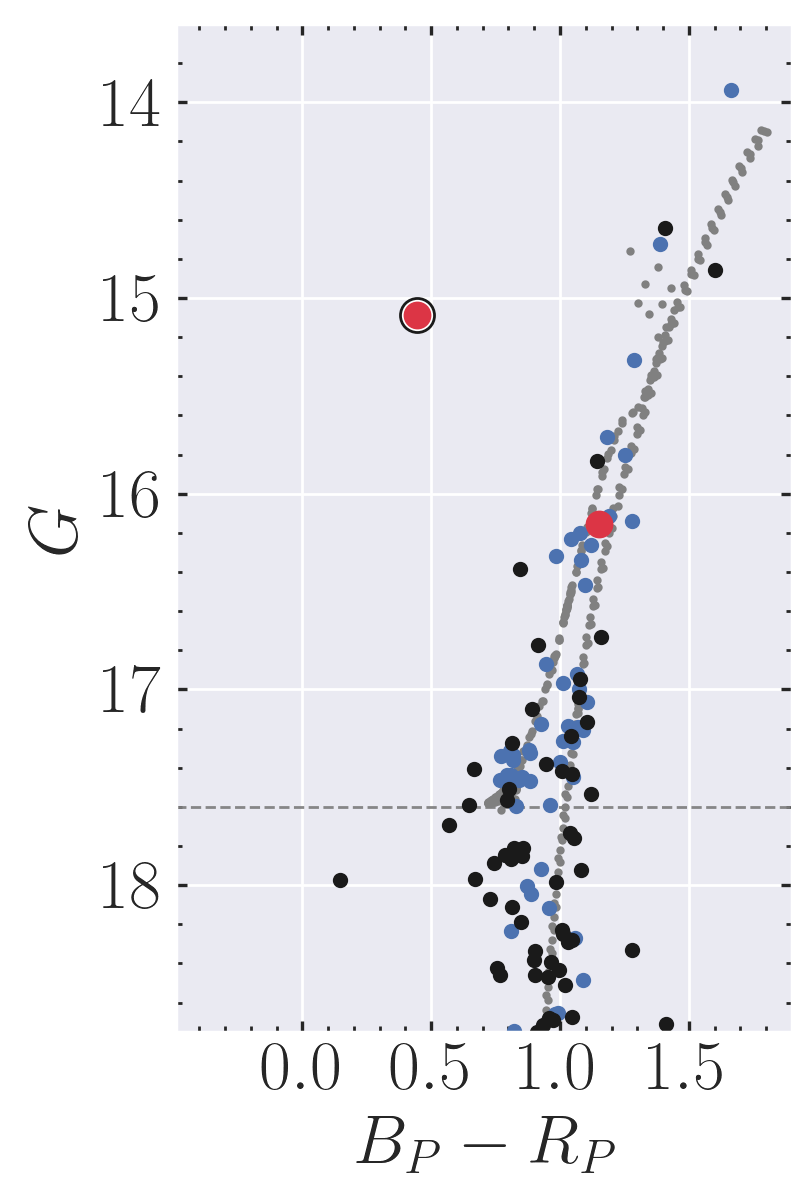}
     \caption{The distribution of the member 
     stars in the color magnitude diagram, 
     $B_P-R_P$ versus $G$ within different 
     longitude $\phi_1$ ranges. The common stars with 
      LAMOST and DESI are marked with red and 
      magenta dots, respectively.
      The horizontal dashed line in each panel
       represents $G=17.6$ for testing 
      the distance gradient.}
    \label{fig:cmd}
\end{figure*}

Among the selected member candidate stars, there are not classical distance tracers, such as 
the RR Lyrae stars or blue horizontal stars. The distance is constrained with the isochrone fitting
in the color-magnitude diagram distribution of the member candidates, as showed in Figure~\ref{fig:cmd}.
Concerning the potential distance gradient, we divide the member candidates into 5 subsamples according
to the longitude $\phi_1$ with binsize of $22^\circ$. The dashed line in each panel represents the 
magnitude in $G-$band of $17.6$, approximately the red horizontal branch stars with distance modulus of
$17.19$. The member stars of the globular cluster Pyxis selected from Gaia DR3 according to the position and
proper motion are also represented by the blue dots and shifted to the similar distance with the stream.

\section{Velocity correction}
\begin{figure}[h]
  \centering
  \includegraphics[width=0.48\textwidth]{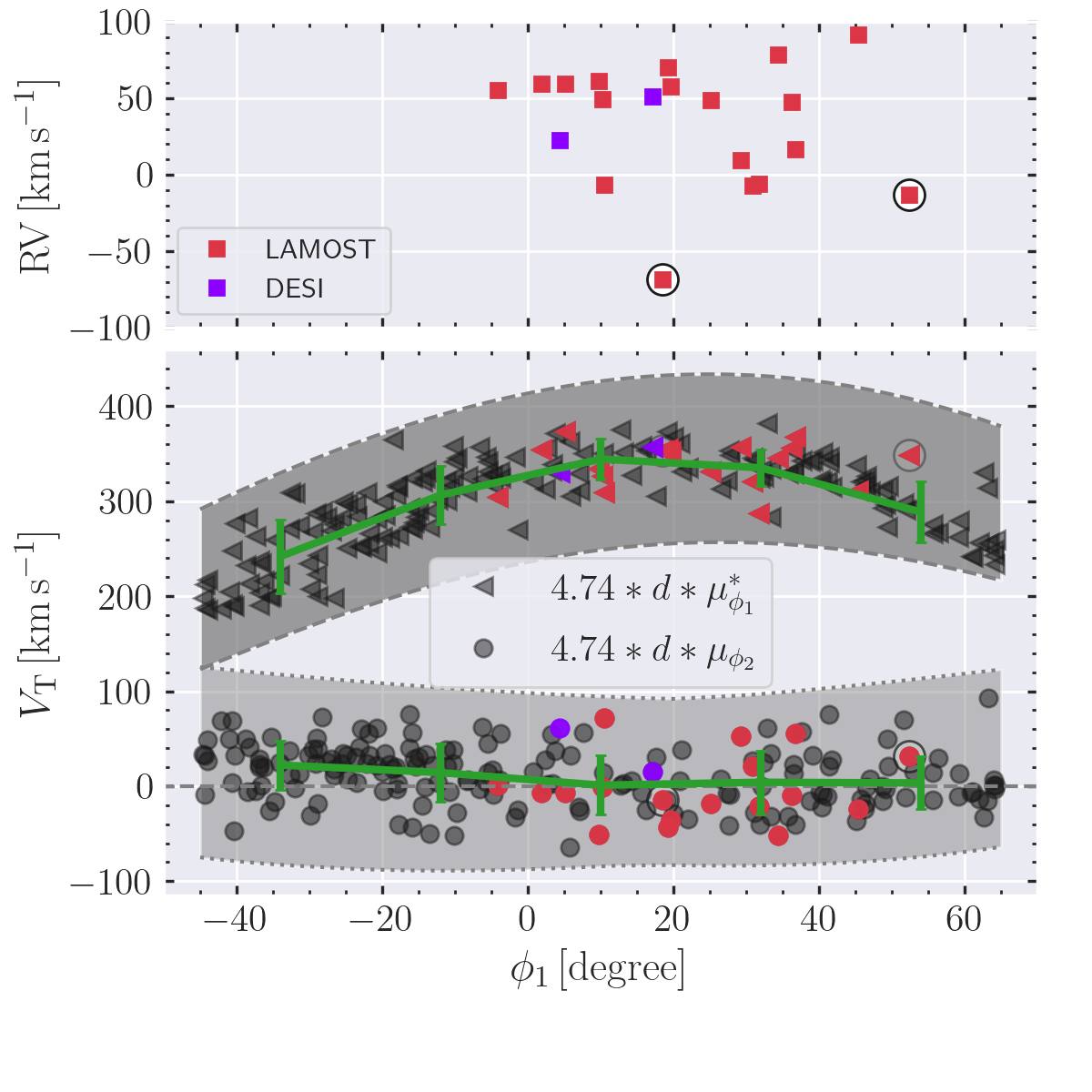}
   \caption{Top: The distribution of the radial velocity $RV$ 
    versus the longtitude $\phi_1$
   of all those common stars from LAMOST and DESI, which are
   represented by the  red and magenta squares, respectively.
  Bottom: The distributions of the 
   tangential velocities $V_T$ versus the longtitude $\phi_1$
   of all the member candidate stars 
   with magnitude $G<18$.  The green 
   lines and the errorbars represent the 
   mean values and the dispersions of the two tangeltial velocities 
   along the two directions $\phi_1$ and $\phi_2$ in each subsample with step
   of $\Delta_{\phi_1}=22^\circ$. The shadow regions represent the edges of the 
   tangential velocities caused by the proper motion selection.
   }
  \label{fig:VVV}
\end{figure}

The contribution of the Solar motion to the proper motions of each stars is different because of 
the different distance and sky position. With distance of each star, this contribution from the Solar motion
can be corrected with \texttt{gala}. The bottom panel in Figure~\ref{fig:VVV} shows the tangential velocities 
versus the longitude $\phi_1$ with the Solar motion corrected. The green symbols represent the average 
values and the dispersion with different longitude $\phi_1$.

\end{appendix}
\end{document}